\documentclass[12pt]{article}
\usepackage{epsfig}

\newcommand{\ba}{\begin{eqnarray}}
\newcommand{\ea}{\end{eqnarray}}

\def\beq{\begin{equation}}
\def\eeq#1{\label{#1}\end{equation}}
\def\eeqn{\end{equation}}

\def\beqa{\begin{eqnarray}}
\def\eeqa#1{\label{#1}\end{eqnarray}}
\def\eeqan{\end{eqnarray}}

\let\bar=\overbar

\def\Dslash{\not{\hbox{\kern-4pt $D$}}}
\def\dslash{\not{\hbox{\kern-2pt $\del$}}}

\def\msb{{\bar{\ssstyle M \kern -1pt S}}}

\def\Title#1{\begin{center} {\Large {\bf #1} } \end{center}}

\begin{document}

\Title{B-meson decay constants and mixing on the lattice}

\bigskip\bigskip


\begin{raggedright}  

{\it Elvira G\'amiz \index{G\'amiz, E.}\\
Department of Theoretical Physics\\
University of Granada / CAFPE\\
E-18071 Granada, SPAIN}
\bigskip\bigskip
\end{raggedright}

\section{Introduction}

It has been argued that differences observed between measurements of some flavor 
observables and the corresponding Standard Model (SM) predictions may be due to beyond the 
SM (BSM) physics affecting the neutral $B$-meson mixing processes~\cite{UTfits1,UTfits1bis}. 
Although the most recent analysis seem to indicate that there are not large BSM contribution to 
neutral $B$-meson mixing~\cite{NiersteCKM12}, the future will bring new twists, and precise calculations of 
the non-perturbative inputs parametrizing the mixing in the SM and beyond are necessary for a thorough 
understanding of quark flavor physics. 
Lattice QCD can provide those non-perturbative theoretical inputs from first principles and with 
errors at the few percent level.

The use of lattice QCD techniques can also shed light on the tension observed between 
the measured value of ${\cal B}r(B\to\tau\nu)$ and SM predictions~\cite{Laiho:2012ss}, by providing precise 
values of $f_B$. Unitarity Triangle fits are very sensitive to $f_B$ and processes with
potential to reveal NP effects depend on $f_B$ or $f_{B_s}$,  so any improvement in the 
decay constant calculations is very important.

In the next Sections I will summarize the status of calculations of both $B$-meson decay constants
and non-perturbative inputs relevant for the study of $B^0-\bar B^0$ mixing using lattice QCD 
techniques. I will only discuss realistic calculations with a complete error budget, simulations 
at several lattice spacings (and an extrapolation to the continuum), and including three dynamical 
quark flavors ($N_f=2+1$ simulations).

\section{Decay constants}

\label{sec:decayconst}

Simulating heavy quarks on the lattice implies having to deal with
discretization errors entering in powers of the mass in lattice units,
$am_Q$. These corrections are not negligible at typical lattice spacings
$a$. In particular, the $b$ quark can not be simulated at its physical mass on present
lattices even with improved actions, since, typically, $am_b>1$. There are two approaches
that have been used to solve this problem:  simulating the heavy quarks with effective theories
(such as heavy-quark effective theories or non-relativistic QCD),
and performing relativistic simulations with improved actions but with smaller masses
than the physical bottom mass (but the same order or
larger than the charm mass) and then extrapolating those results up
to the physical $m_b$. The first approach is being used by the Alpha (heavy-quark
effective theory, HQET), HPQCD (non-relativistic QCD, NRQCD), FNAL/MILC (Fermilab action),
and RBC/UKQCD (non-peturbatively relativistic heavy-quark action) Collaborations.
One of the dominant systematic errors in these calculations is the one associated with the use
of an effective theory. The collaborations that follow a relativistic
approach for $b$ quarks and whose results I will mention here are ETMC (tm action) and
HPQCD (HISQ action).

There have been three lattice $N_f=2+1$ calculations of $f_B$ and $f_{B_s}$ in the
last two years by the HPQCD~\cite{HPQCDfB1,HPQCDfB2} (with a relativistic and a 
non-relativistic approach respectively) and the FNAL/MILC~\cite{FMILCfB}
collaborations, which have reduced the error to the $2.5\%$ level. The smallest error 
in \cite{HPQCDfB2} for $f_B$ is achieved by determining the ratio $f_{B_s}/f_B$ using a
non-relativistic description (NRQCD) of the $b$ quarks together with the $f_{B_s}$
determination in \cite{HPQCDfB1}, which employs relativistic $b$ quarks.
By using the ratio $f_{B_s}/f_B$ the dominant systematic errors associated with the
effective NRQCD description partially cancel, so this determination is nearly
free of uncertainties due to an effective description of the $b$ quark.
The average values of $f_B$ and $f_{B_s}$ from these three calculations are~\cite{LLV}
\ba\label{eq:fB}
\hspace*{-0.5cm}f_B = 190.6(4.7){\rm MeV}\,, \,\,\,
f_{B_s} = 227.6(5.0){\rm MeV}\,,\,\,\,{\rm and}\,\,\,\frac{f_{B_s}}{f_{B_d}}=1.201(17)\,.
\ea

The direct comparison of the results in Eq.~(\ref{eq:fB}) with experiment 
is problematic due to the need of the value of the CKM matrix element $\vert V_{ub}\vert$ 
(whose inclusive and exclusive determinations disagree at the $3\sigma$ level) and the 
$\sim 2\sigma$ disagreement of BaBar~\cite{BaBarBtolnu} and Belle's~\cite{BelleBtolnu} measurements. 
Nevertheless, Belle new measurement seems to alleviate the tension between theory and experiment. 

In the next two years there will be new results for $B-$meson
decay constants with $N_f=2+1$ and $N_f=2+1+1$ from lattice collaborations using
both an effective theory description (RBC/UKQCD~\cite{RBCBphys}) and a
relativistic description (ETMC~\footnote{ Their $N_f=2$
results~\cite{ETMCfB} agree within $1\sigma$ with Eq.~(\ref{eq:fB}).}),
as well updates with considerably smaller errors from FNAL/MILC~\cite{lat11decayconst}.

\subsection{Neutral $B-$meson mixing}

\label{sec:Bmixing}

The current status of $N_f=2+1$  lattice calculations of the non-perturbative quantities
parametrizing the mass differences between the heavy and the light
mass eigenstates in both the $B^0_d$ and $B^0_s$ systems, as well as the SU(3) breaking
ratio $\xi$, is summarized in Tab.~\ref{tab:mixing}. The HPQCD~\cite{BmixingHPQCD}
and FNAL/MILC~\cite{BmixingFMILC} collaborations
use the same light quark formulation, but a different description for the $b$ quarks.
The exploratory study by the RBC/UKQCD collaboration uses heavy quarks in the static 
limit~\cite{BmixingRBC}.
The average of the results in Tab.~\ref{tab:mixing} for $\xi$
gives the value $\xi=1.251\pm0.032$. The calculations whose results are shown in 
Tab.~\ref{tab:mixing} were not optimized to extract the bag parameters but rather the matrix 
elements, so the errors for the bag parameters can be significantly reduced to about 
a 3\% error in on-going calculations.

\begin{table}[t]
\begin{center}
\begin{tabular}{cccc}
\hline
\hline
 & HPQCD & FNAL/MILC & RBC/UKQCD \\
\hline
$\xi$ & 1.258(33) & 1.27(6) & 1.13(12) \\
\hline
$B_{B_s}/B_{B_d}$ & 1.05(7) & 1.06(11) & - \\
\multicolumn{4}{c}{HPQCD: $f_{B_s}\sqrt{\hat B_{B_s}} = 266(6)(17)~{\rm MeV}$,\hspace*{0.4cm}
$\hat B_{B_s}=1.33(6)$}\\
\multicolumn{4}{c}{HPQCD: $f_{B_d}\sqrt{\hat B_{B_d}} = 216(9)(13)~{\rm MeV}$,\hspace*{0.4cm}
$\hat B_{B_d}=1.26(11)$}\\
\hline
\hline
\end{tabular}

\end{center}

\vspace*{-0.3cm}
\caption{$B-$meson mixing parameters. $\xi$ is defined as the ratio of the
parameters in the second and third rows. In the case where there are two errors,
the first one is statistical and the second one systematic. \label{tab:mixing}}
\end{table}
There is not yet a finalized calculation of the matrix elements needed for the determination of
the decay width differences, $\Delta\Gamma_{d,s}$, in the continuum limit and with
$N_f=2+1$ flavors of sea quarks, but preliminary results for the relevant matrix elements
by FNAL/MILC can be found in~\cite{Bmixinglatt2011}.

Beyond the SM the mixing parameters can have contributions from $\Delta B=2$
four-fermion operators which do not contribute in the SM. The matrix elements of
the five operators in the complete basis describing
$\Delta B=2$ processes, together with the Wilson coefficients for those operators
calculated in a particular BSM theory and the experimental measurements of the mixing
parameters, can provide very useful constraints on that BSM theory. Again, there is not a final
unquenched lattice calculation of the matrix elements
of all the operators in the $\Delta B=2$ effective hamiltonian, but FNAL/MILC presented
preliminary results for the complete basis in~\cite{Bmixinglatt2011}.

The authors of Ref.~\cite{BurasBmumu} suggested that the branching fractions of the rare
decays $B_q\to \mu^+\mu^-$ (for $q=s,d$) could be determined from the experimental
measurement of the mass difference in the neutral $B_q$-meson system, $\Delta M_q$, and
the lattice calculation of the bag parameter $\hat B_{B_q}$ using
\ba\label{eq:BagBmumu}
\frac{{\cal B}r ( B_q \to \mu^+ \mu^- )}{\Delta M_q}
= \tau(B_q)\,6\pi                                                                                              
\frac{\eta_Y}{\eta_B}\left(\frac{\alpha}{4\pi M_W sin^2\theta_W}\right)^{2}                                    
\,m_\mu^2\,\frac{Y^2(x_t)}{S(x_t)}\,    
\frac{1}{\hat B_q}\,.
\ea
In order to compare experimental measurements and the theory predictions for the decay
rate of $B^0_s$, one must include the effects of a non-vanishing
$\Delta\Gamma_s$~\cite{BsmumuFleischer}. This can be done in the SM by rescaling the
theory prediction by $1/(1-y_s)$, where
$y_s\equiv\tau_{B_s}\Delta\Gamma_s/2$~\cite{BsmumuFleischer}.
Multiplying Eq.~(\ref{eq:BagBmumu}) by this factor for the $B^0_s\to\mu^+\mu^-$ decay
and using the HPQCD determination of the bag
parameters $\hat B_{B_s}=1.33(6)$ and
$\hat B_{B_d}=1.26(11)$~\cite{BmixingHPQCD}; together with $\tau_{B_s}=1.497(15){\rm ps}$,
$\tau_{B_d}=1.519(7){\rm ps}$~\cite{PDG12}, and  $\Delta\Gamma_s = 
0.116(19){\rm ps}^{-1}$~\cite{LHCbDeltaGamma}, one gets
\ba\label{eq:Btomumu}
{\cal B}r
( B_s \to \mu^+ \mu^- )\vert_{y_s} & = & (3.65\pm0.20)\times 10^{-9}\,,\nonumber\\
{\cal B}r ( B_d \to \mu^+ \mu^- ) & = & (1.04\pm0.09)\times 10^{-10}\,.
\ea

The direct calculation of these branching fractions has become competitive with the one
in (\ref{eq:Btomumu})~\cite{BurasGirrbach} thanks to the recent improvements in the
calculation of the $B$-meson decay constants on the lattice summarized in
Sec.~\ref{sec:decayconst}.
Including the correction factor $1/(1-y_s)$ for the $B_s\to\mu^+\mu^-$ decay rate,
and using the same inputs as in Ref.~\cite{BurasGirrbach}
except for $f_B$ and $f_{B_s}$, for which I use the averages described in
Sec.~\ref{sec:decayconst}, and $\tau_{B_s}$, which I take
equal to its PDG 2012 value, $\tau_{B_s}=1.497(15){\rm ps}$, I get

\ba\label{eq:Btomumudirect}
{\cal B}r ( B_s \to \mu^+ \mu^- )\vert_{y_s} & = & (3.64\pm0.23)\cdot 10^{-9}\,,\nonumber\\
{\cal B}r ( B_d \to \mu^+ \mu^- ) & = & (1.07\pm0.10)\cdot 10^{-10}\,.
\ea
The agreement between the two set of numbers in (\ref{eq:Btomumu}) and
(\ref{eq:Btomumudirect}) is excellent. This gives us confidence in the SM prediction for
these branching fractions, and this confidence will increase when we have results
for the bag parameters entering in (\ref{eq:Btomumu}) from the on-going lattice calculations
described above. This is very important since the LHC bounds are now approaching the SM 
predictions~\footnote{After this conference, the LHCb Collaboration announced the ﬁrst 
evidence for one of these two processes~\cite{LHCb:2012ct}, ${\cal B}r ( B_s \to \mu^+ \mu^- )= 
\left(3.2^{+1.5}_{-1.2}\right)\cdot 10^{-9}$. 
This result agrees with the SM prediction given in this proceeding.}, 
especially for $B^0_s$ decays, ${\cal B}r ( B_s \to \mu^+ \mu^- )\vert_{LHC}
< 4.2\cdot 10^{-9}$ at 95\% CL~\cite{LHCboundsBmumu}.

\section{Outlook and future prospects}

In the near future there will be results for the $B$-meson mixing parameters and the decay constants 
at the percent level and with different lattice formulations for both light and heavy quarks. 
Some of the results expected for next year are: the first unquenched 
determinations in the continuum limit of matrix elements needed for the calculation of 
the decay width differences $\Delta\Gamma_{s,d}$, 
the analysis of $B$ mixing in BSM theories, and the study of 
short-distance contributions to $D^0-\bar D^0$ mixing.

Further improvement will be achieved by using simulations at the physical $u$ and $d$ masses 
(some preliminary results for $f_B$ are already available from the HPQCD collaboration), 
including the effect of dynamical charm quarks, and extending the use of relativistic actions 
to the analysis of $B^0-\bar B^0$ mixing. 

\section{Acknowledgements}
The author's work is supported in part by the MICINN under Grant FPA2010-16696 and
\emph{Ram\'on y Cajal} program; by Junta de Andaluc\'{\i}a under Grants FQM-101, FQM-330,
and FQM-6552 (E.G.); and by European Commission under Grant No.
PCIG10-GA-2011-303781.

\def\Discussion{
\setlength{\parskip}{0.3cm}\setlength{\parindent}{0.0cm}
     \bigskip\bigskip      {\Large {\bf Discussion}} \bigskip}
\def\speaker#1{{\bf #1:}\ }
\def\endDiscussion{}

\end{document}